\documentclass{aa}

\begin{document}

\author{Biping Gong\inst{1,2}, G.F. Bignami\inst{1}}
\institute{Centre d'Etude Spatiale des Rayonnements, CNRS-UPS, 
9, avenue du Colonel Roche, 31028 Toulouse Cedex 4, France
\and
Department of Astronomy, Nanjing University, Nanjing
210093, P.R. China}

\offprints{Biping Gong  \email{bpgong@nju.edu.cn} }
\authorrunning{Gong $\&$ Bignami}

\date{Received  / Accepted }


\title{
On the nature  of  the radio quiet X-ray neutron star 1E\,1207.4-5209
}

\titlerunning{On the nature of 1E1207}

\abstract{
The strange timing property of X-ray pulsar 1E\,1207.4-5209 can be explained 
by the hypothesis that it is  a member of an ultra-compact binary system. This 
paper confronts  the ultra-compact assumption with the observed 
properties of this pulsar. The gravitational potential 
well of an ultra-compact binary can enlarge the corotation radius and 
thus make it possible for accreting material to reach the surface  
of the NS  in the low accretion rate case. Thus 
the generation of the absorption features should be similar to the 
case of accreting pulsars.
The close equality of  the energy loss by 
fast cooling of the postsupernova neutron star and  the 
energy  dissipation needed for a wide binary evolving to an ultra-compact 
binary demonstrates that the ultra-compact binary may be formed in 10-100yr 
after the second supernova explosion.  
Moreover, the ultra-compact binary hypothesis can well explain the 
the absence of optical counterpart and the observed  two black body 
emissions.
We suggest a simple method which  can test the binary nature directly 
with  XMM-Newton  and Chandra observations. We further predict that 
the temperature of the two black bodies 
should vary at different pulse periods.  
}

\maketitle

\section{Introduction}
The X-ray pulsar 1E\,1207.4-5209 (hereafter 1E1207) is an very 
strange object. The X-ray emission with $P=0.4241296$s
period~(\cite{Zavlin00}) indicates that is a neutron star (NS). But 
there is no radio, no gamma-ray and no optical emission detected.     
The association with the surrounding supernova remnant (SNR) PKS 
1209-51/52 is very likely~(\cite{Pavlov02}), however the age of SNR is 
substantially different from that of the spin-down of the pulsar. 
The  X-ray features show cyclotron absorption, from which  the B-field can 
be obtained. This is very different from the magnetic field 
(B-field) inferred 
from  the  spin-down of the pulsar~(\cite{Bignami03,Zavlin04}). 
Moreover, the pulse frequency variation is 
non-monotonic~(\cite{Zavlin04}). 
These puzzles are difficult to 
explain by  a normal, isolated pulsar.   

Can it be an isolated pulsar with an accretion disk? 
Supernova explosion  ejecta may form a disk around it. 
The disk  exerts a torque 
on the NS and makes it spin down. This model can explain the age 
and possibly the B-field problem.
However,  the pulse frequency is non-monotonic,  
which implies a 
variable torque,  and hence a variable flux 
density. Moreover, the jump of pulse frequency leads to a high 
luminosity ($\sim 10^{36}$erg s$^{-1}$). But  the 
observed luminosity is stable, around 
$\sim 10^{33}$erg s$^{-1}$. 
Therefore,  accretion    as 
the major cause of the pulse frequency variation is not suggested by 
 observations~(\cite{Zavlin04}).

If the pulse frequency variation is interpreted as due to one or more 
glitches,  then 
the amplitude of the glitch of 1E1207  ($\Delta\nu>5\mu$Hz) is a factor of 30-60 
greater than 
the typical value observed for glitching radio pulsars, which is 
unlikely~(\cite{Zavlin04}).

Can it be a  binary system? The binary hypothesis with orbital 
period from 0.2yr to 6yr can explain the variation  of pulse 
frequency~(\cite{Zavlin04}). However this  implies that the variation of the pulse 
frequency must be 
periodic, which needs to be confirmed by further observations. 
On the other hand,   the wide binary hypothesis can not explain other puzzles, 
i.e., age and  B-field values. 
Moreover, repeated optical observations have failed to yield any 
optical counterpart for the system~(\cite{Luca04}).

Can it be an ultra-compact binary? This model may 
explain  the pulse frequency variation by a long-term orbital effect,  
 not strictly periodic. Moreover,   the age 
puzzle and the B-field puzzle can also be explained~(\cite{Gong05}). 
This paper  confronts  the  ultra-compact binary model with 
all  available observations.

We demonstrate  that the ultra-compact binary system shares a common 
accretion disk, from which the generation of cyclotron absorption 
features is naturally expected.

A new scenario for the formation of an ultra-compact binary is 
proposed, in which  an ultra-compact binary can be formed from 
a wide presupernova 
binary through  energy loss by  fast cooling in 10-100 yr.

Why the orbital motion of 1E1207 has not been 
measured is also analyzed in detail.   A  simple method  to test  the binary 
nature of an ultra-compact binary candidate using 
XMM-Newton and Chandra data is proposed.

We further predict that the temperatures of the two black bodies 
should vary at different pulse periods.



\section{The search for an  orbital period in 1E1207}

This section analyzes  why the orbital period of 1E1207  can escape 
different tests, like modulation of flux density, Doppler shift of 
pulse frequency, and Roemer delay,  and how to  test the  binary nature by
XMM-Newton and Chandra data in an alternative way. 


In the case of a circular orbit, the phase delays caused by the motion 
of the pulsar are sinusoidal. The signal received by a telescope can 
be described by~(\cite{Ransom03})  
\begin{equation}
\label{ransom} f(u) =A\cos [\frac{2\pi t}{P}
+\phi_{\rm spin}+\Phi_{\rm orb}\cos(\frac{2\pi t}{P_{\rm b}}+\phi_{\rm orb})]
  \,,
\end{equation}
where $A$ and $\phi_{\rm spin}$ are the amplitude and phase  of the 
pulsation,
$\Phi_{\rm orb}\equiv{2\pi x}/{P}$ and $\phi_{\rm orb}$ are amplitude and phase of the 
orbital modulation, $P$, $P_{\rm b}$ and  $x$ are pulse period, 
orbital period, and projected semi-major axis respectively. 
The  binary nature of 1E1207 can be tested in four different ways.

(1) 
As in radio pulsars, the  binary nature can be tested by the Roemer 
time delay, which is the propagation time in the orbit. However, the 
short orbital period $P_{\rm b}$ corresponds to the 
small projected semi-major axis, $x$ ($x\equiv a_{1}sin 
i/c\sim 3$ms), which is 
comparable or even smaller than the timing resolution of 
XMM-Newton, $\sim$6ms. Thus the binary nature of 1E1207 
cannot be tested directly by the time of arrivals~(\cite{Gong05}).

(2)
If 1E1207 is in  an ultra-compact binary system, then the measured 
photon time 
series should show evidence for orbital modulation. However ,
this has not been measured. The reason is that 
the largest modulation which corresponds to the pulsation has an 
amplitude of about $6\%$~(\cite{Luca04}). 
Consequently, the amplitude of the orbital modulation, 
which is two to three orders of magnitude smaller 
($\Phi_{\rm orb}\sim 10^{-2}-10^{-3}$) than  $6\%$, cannot be 
observed, as shown in Eq.~($\ref{ransom}$).

(3) 
The  Fourier response of the fundamental spin harmonic corresponding 
to the signal of 
Eq.~($\ref{ransom}$) is given by Ransom et al (2003). Since the  
amplitude and the 
separation of the sideband are both dependent of  $\Phi_{\rm orb}$, which 
are small for  ultra-compact binaries (due to $x\sim 10^{-2}-10^{-3}$s), 
 the side bands due to orbital motion are very difficult to resolve 
 from the noise. Thus also this method is not suitable in searching 
 ultra-compact binaries with orbital periods of a few minutes.

(4)
The last possibility is  measuring the variation of  pulse 
frequency, i.e., analyzing $\nu$ in the whole time span
(typically $\tau \sim 100$ks). This is efficient 
for binary pulsars 
with wide orbits, in which $P_{\rm b}\sim\tau $ or  $P_{\rm b}>\tau$. 
If the orbital period is of a few minutes, then $P_{\rm b}<<\tau$ and  
 the measured $\nu$ is affected only  by the long-term orbital
effect, 
\begin{equation} \label{av2}
\Delta\nu  ={x\kappa\nu}\,\pi\,(1-{e^2}/{4})/{P_{\rm b}}
+o(\frac{P_{\rm b}}{\tau})
 \ .
\end{equation}
The second term in the right hand side of  Eq.~($\ref{av2}$) is  
$P_{\rm b}/{\tau }$ ($\sim 10^{-3}-10^{-4}$)
times smaller than  the first  and can be neglected. The 
 $\Delta\nu$ of Eq.~($\ref{av2}$) depends on the 
orbital elements, $a$, $i$ (which are contained in $x$), $e$ and 
$P_{\rm b}$, which are long-periodic terms due to spin-orbit coupling~(\cite{Gong05b}).   
However, these long-periodic terms are not exactly periodic, 
implying that  the variation of $\nu$ is also  not periodic.  
This  may be responsible for the side band on the  probability 
density distribution of $\nu$~(\cite{Zavlin04}).

Consequently, analyzing $\nu$ in the whole observational time span
can only reflect the long-periodic property of an ultra-compact binary 
system (at time scales $\sim 10^{4}P_{\rm b}$), 
but not the short-period effect (at time scales $\sim P_{\rm b}$). 

Thus, it is the small value of the orbital period of an ultra-compact binary  that 
prevents the measurement of the orbital motion of 1E1207 directly. 
However, the two NSs in an ultra-compact binary have  
large orbital velocities, $\sim 10^{3}$km s$^{-1}$, 
which means that the Doppler shifts can be as large as   $v/c\sim 10^{-2}-10^{-3}$. 
Using  the large Doppler shifts it is still possible to 
extract the orbital period of 1E1207 directly. 

However, a large Doppler shift of 1E1207 can only 
be found at $\sim P_{\rm b}/2$ time scale, or around 1 minute.
If one  measures  $\nu$  at the time scale of 1 
minute directly, then the signal to noise ratio is far too  small.

A simple solution to the problem  uses the large Doppler 
shift and the length of time span ($\tau>>P_{\rm b}$) simultaneously.   
One can split the time span evenly into  a 
number of segments, and each segment corresponds to a time 
scale of $\sim P_{\rm b}/2$.  Then fold the odd segments, 
which may correspond to one shift (say blue); and fold separately the even 
segments, which may correspond to the other shift (red).   
Thus the orbital period can 
be tested by  comparing the shifts in the two groups of folding.  

Assume that the total time span of an observation is $\tau\sim 
100$ks, and that the folding starts at  i.e., $t_{\rm 0}=70$s. Thus, the total 
of folding time is $\tau-t_{\rm 0}$, which can be divided into (for simplicity) $6$ 
segments, No 1,2,3$\ldots$6,
and each segment corresponds to a sub-time scale of 
$\tau_{\rm i}=N\cdot P=42.4$s ($N=100$ denotes the number of pulses, 
recall $P=0.42$s is the pulse period). 
Then we can fold the two groups of odd and even segments separately.
In such a case, the total time span of an observation is 
$\tau=t_{\rm 0}+6\tau_{\rm i}$.

Having  the two groups of 
folding, one can test the binary nature in 
two ways.

(a) Using the method of Zavlin et al. (2004),  the $\nu$ and  
$\dot{\nu}$ values of the two groups of folding can be obtained. Due to the orbital 
motion, $\nu$ may be greater than a reference central frequency,  
$\nu_{\rm ref}$; and $\dot{\nu}>0$ in the folding 
corresponding to blue shift,  whereas in the folding 
corresponding to red shift $\nu$ may less than $\nu_{\rm ref}$ and $\dot{\nu}<0$. 
 
(b) Since the Doppler shift can be as large as $\Delta\nu/\nu\sim 
1\%$, then the  $\nu$ of the blue folding is larger than that of red 
folding  by about  $1\%$. Thus, one can find one more pulsation in 
the blue folding than that in red folding in the given time scale, 
$P_{\rm b}/2$, when the number of pulsations in the interval $P_{\rm b}/2$ is 
about 100.

Notice that, in the performance of the two above tests 
one can  change the length of each segment, $\tau_{\rm i}$ (through changing $N$), 
which corresponds to the orbital period 
of the binary, and the initial time, $t_{\rm 0}$, of the segments which 
corresponds to the initial orbital phase of the binary,
in order to obtain the maximum effect of (a) and (b).

It is conceivable that the measured pulse profile of 1E1207 
may have  been broadened by the orbital 
Doppler shift. By excluding this effect, more precise pulse 
profiles  are expected.

\section{Accretion disk of an ultra-compact binary}
\label{sec:disk}

In this section we demonstrate that 
the accretion disk of an ultra-compact binary is different  
from that of other compact systems in two aspects.

One is the corotation radius, $r_{\rm c}$, which is  enlarged by the 
potential well of the ultra-compact binary. Thus  the accretion 
regime can be reached even in the low accretion rate case.  

The other is the inner radius of the disk, which is larger than the 
separation of the two compact stars, so that the two 
compact stars 
share one common accretion disk.

The accretion can proceed in two regimes~(\cite{King, Zavlin04}), depending on the 
relation between the corotation radius, $r_{\rm c}$ and the magnetospheric 
radius, $r_{\rm m}$. The former is given by,  
\begin{equation}
\label{rc} r_{\rm c}=(Gm_{1})^{1/3}(2\pi \nu)^{-2/3}=0.1\times 10^{9} cm \,,
\end{equation}
(where $\nu=2.36$Hz is the spin frequency, $m_{1}$ is the mass of the NS) 
and the latter is given by, 
\begin{equation}
\label{rm} r_{\rm m}\sim 0.5(8Gm_{1})^{-1/7}\mu^{4/7} \dot{m}^{-2/7}
\approx 0.9\times 10^{9} \mu_{30}^{4/7} \dot{m}_{14}^{-2/7}cm \,,
\end{equation}
where $\dot{m}=10^{14}\dot{m}_{14}$g s$^{-1}$ is the accretion rate,  
$\mu=10^{30}\mu_{30}$G cm$^{3}$ is the magnetic moment, and the NS 
mass is assumed, $m_{1}=1.4m_{\odot}$.

If the accretion rate is high enough,   $r_{\rm c}>r_{\rm m}$, the accretion 
matter can overcome the centrifugal barrier and reach the NS surface. 
In this case the torque exerted on the magnetosphere spins the NS up. 
When $r_{\rm c}<r_{\rm m}$, the centrifugal force at $r=r_{\rm m}$ exceeds the 
gravitational force, so that accretion on to the NS surface is 
inhibited.

By Eq.~($\ref{rc}$) and Eq.~($\ref{rm}$), we have $r_{\rm c}<r_{\rm 
m}$, which corresponds to a  propeller regime  in the case of an 
accretion disk around an isolated pulsar. Transforming to the accretion 
regime,  $r_{\rm c}>r_{\rm m}$, needs to substantially increase the $\dot{m}$ 
and hence X-ray luminosity. This is not supported  
by observations. Thus the variation of spin frequency 
is not likely to be due to accretion.

On the other hand, the pulse frequency  variation can be interpreted 
as  the dynamic effect of an 
ultra-compact binary~(\cite{Gong05}). If this is true, then the  
question is:  how does the accretion disk interact with the ultra-compact binary,  
and what is the difference between  a disk in such case and that  
of  normal X-ray binaries. The potential of a binary system is given,  
\begin{equation}
\label{phr} 
\Phi_{\rm R}({\bf r})=-\frac{Gm_{1}}{|{\bf r}-{\bf r}_{1}|}
-\frac{Gm_{2}}{|{\bf r}-{\bf r}_{2}|}
-\frac{1}{2}({\bf \omega}_{\rm b}\times {\bf r})^{2}
\,,
\end{equation}
where ${\bf r}_{1}$ and  ${\bf r}_{2}$ are position vectors of 
the centers of the two stars,  ${\bf r}$ is the position of a field point 
from the common center of mass, and ${\bf \omega}_{\rm b}\equiv 2\pi/P_{\rm b}$ is the angular 
velocity of the binary system. 
The Roche lobe radius, $R_{\rm L}$, is given~(\cite{King})
\begin{equation}
\label{lobe} 
\frac{R_{\rm L}}{a}=[0.38-0.2\log(\frac{m_{2}}{m_{1}})], Ê\ 
\,  0.05<\frac{m_{2}}{m_{1}}<2 \ .
\end{equation}
In the case ${m_{2}}/{m_{1}}\approx 1$ and $P_{\rm b}=1$min, we have 
$R_{\rm L}=0.38a=0.8\times 10^{9}$cm.
 Eq.~($\ref{rm}$) gives a magnetosphere radius, $r_{\rm m}\sim 
0.9\times 10^{9}$cm. Thus $r_{\rm m}>R_{\rm L}$,  which implies that 
 a disk inside 
the Roche lobe of 1E1207 (say primary for convenience) is impossible.
In such a case, the only possibility is that  the  inner edge radius of 
the disk is larger than the separation of the 
two compact stars. 

For a given angular momentum, a circular orbit has the minimum  energy, so 
dissipation in the orbiting gas will tend to circularize the motion.   
If the inner edge of the disk is 
approximately circular, then the Roche potential of Eq.~($\ref{phr}$), 
can be denoted by an equivalent
 potential, $\Phi\equiv-{GM}/{r}$ ($M=m_{1}+m_{2}$),
\begin{equation}
\label{Phi2} 
\Phi\equiv-\frac{GM}{r}\approx\Phi_{R} \ .
\end{equation}
Then we have, 
\begin{equation}
\label{Phi3} 
\frac{|\Phi-\Phi_{R}|}{|\Phi|}<\frac{1}{10}, \ \ \ \  r>5a \ ,
\end{equation}
where $5a\approx 1\times 10^{10}$.
In the case of Eq.~($\ref{Phi3}$),  the 
gravitational force exerted on the inner edge of the disk equals 
approximately  that of  a star with mass $M$. 

Therefore, 1E1207 and its companion may share an accretion disk, 
from which material can fall onto both compact stars.
The compact system is thus different from a normal X-ray binary, which is characterized by 
a  much 
larger separation of the two stars and whose disk is in the 
Roche lob of the primary.

The potential well of a pulsar in a binary system,  given by Eq.~($\ref{phr}$), 
 is deeper than that of an isolated pulsar. 
Consequently,  
a pulsar in an ultra-compact binary system has a larger 
corotation radius, $r_{c}$, than that of an isolated pulsar.

To obtain the effective corotation radius, it is sufficient  to 
calculate it in the simple case when   
${\bf r}$, ${\bf r}_{1}$ and  ${\bf r}_{2}$ are aligned. 

Assuming $\Phi_{\rm R}(r_{\rm eff})=-Gm_{1}/r_{\rm c}$, the effective 
radius,  $r_{\rm eff}$, can be  obtained.  
When  $P_{\rm b}=1$min and $P_{\rm b}=3.3$min, we 
have $r_{\rm eff}\approx 4r_{\rm c} $ and  $r_{\rm eff}\approx 7r_{\rm c}$ respectively. 
Thus the potential of 
Eq.~($\ref{phr}$)  corresponds to a larger  
corotation radius, and hence,  even in the low accretion rate  case, i.e.,  
$\dot{m}\sim 10^{14}$g s$^{-1}$,  it is still possible that the relation 
$r_{\rm c}>r_{\rm m}$ be satisfied, and the pulsar is in  
 the accretion regime.  On the contrary, if 1E1207 were to be an 
 isolated pulsar, the  accretion rate in 
 $\dot{m}\sim 10^{14}$g s$^{-1}$ would correspond to the propeller regime, 
 as shown in Eq.~($\ref{rc}$) and Eq.~($\ref{rm}$).

It is worth mentioning that the influence of the B-field of the 
secondary  on the magnetic moment of the primary,
$\mu=BR^3$, is negligible.
 Because we do not need to consider the B-field between the two compact stars, 
and in the region  outside of the two  stars, the B-field is dominated by each 
star's own B-field (since the B-field of the secondary near  the 
surface of the primary is $B\sim (R_{*}/a)^{3}B_{2}\sim 10^{-9}B_{2}$, with 
$R_{*}$ the 
radius of the primary, and $B_{2}$ the B-field of the secondary). 

Consequently, the ultra-compact binary scenario can lead to a significant increase  
in the corotation radius of each NS, while causing insignificant changes in 
their magnetospheric radius.

\section{The formation of an ultra-compact binary}
The age of 1E1207 should be that of that of SNR hosting it, which is about 
$3-20$kyr~(\cite{Roger88}). But forming an ultra-compact binary through gravitational 
wave radiation takes a much longer time span. This problem can be 
solved by a simple scenario. 
We assume that the presupernova 
binary is a wide one, which makes it possible for it  to survive the second 
SN explosion, whose remnant we  observe. The postsupernova binary
(still with a wide orbit) loses 
an enormous amount of energy in 10-100yr, through fast cooling.

The energy released in the fast cooling corresponds to a pulsar wind 
of $(10^{-5}M_{\odot}-10^{-6}M_{\odot})$yr$^{-1}$, which interacts 
with the SN envelope  and reduces the mechanic  
energy of the binary system rapidly.

On the other hand, the energy released in the fast cooling  is 
approximately the  energy that is required for a  wide binary to decay  to 
an ultra-compact one.  This implies that  fast cooling may be  the 
cause of formation an ultra-compact binary system in a relatively 
short time scale.

NSs are formed at very high temperatures, $\sim 10^{11}$K, in the 
imploding cores of supernova explosions. Much of the initial thermal 
energy is radiated away from the interior of the star by various 
processes of neutrino emission,  
leaving a one-day-old NS with an internal temperature of 
about   $\sim 10^{9}-10^{10}$K~(\cite{Becker02}) .

The relationship between surface temperature, $T_{\rm s}$ 
and the core temperature $T_{\rm i}$ of a NS is given~(\cite{Goodman83})
\begin{equation}
\label{Ts} T_{\rm s}=3.1(g/10^{14}cm s^{-1})^{1/4}(T_{\rm i}/10^{9})^{0.549} \ \ K
\end{equation}
where g is the gravitational acceleration at the NS surface.

Having cooled down to $T_{\rm s}=1.5-3$MK (1MK=$1\times 10^{6}$K), 
the NS surface temperature stays 
on a plateau for several decades. 
The cooling can then follow  two different 
scenarios, depending on the still poorly known properties of 
super-dense matter and on the mass of NS~(\cite{Page92}).

For a NS with  mass, $m\leq  1.3m_{\odot}$, the cooling is not by direct 
Urca process~(\cite{Page92}) and  the temperature decreases gradually, down 
to  $\sim 0.3-1$MK, by the end of the neutrino cooling era. It then 
falls down  exponentially, becoming lower than  $\sim 0.1$MK in  $\sim 
10^{7}$yr~(\cite{Becker02}).

For  NS with  mass, $m\geq   1.35m_{\odot}$, the interior of a star  cools very rapidly
 by the direct Urca process. A sharp drop 
in temperature, down to 0.3-0.5MK, occurs at an age of  $\sim 
10-100$yr, followed by a more gradual decrease, down to the same 
$\sim 0.1$MK at  $\sim 10^{7}$yr~(\cite{Page92, 
Lattimer91,Becker02,Vakov04}).    

From  Eq.~($\ref{Ts}$), the decrease of the surface temperature from $T_{\rm s}=1.5-3$MK, 
to 0.3-0.5MK corresponds to a decrease of core
temperature,  
\begin{equation}
\label{DeltaT} \Delta T_{\rm i}\sim 1\times 10^{9} \ \ K  \ \ . 
\end{equation}
This temperature drop corresponds to an energy dissipation of  
\begin{equation}
\label{DeltaE1} \Delta E_{1}\sim \frac{3}{2}k \Delta T_{\rm i}\sigma N    \ \ , 
\end{equation}
where $N$ denotes the total number of nuclei in a compact star, 
and $\sigma$ ($0<\sigma< 1$) denotes the ratio of the  
number of nuclei contained in the 
core to that of the whole NS.

For each proton-electron pair accreted, the potential energy released 
is  $GM(m_{\rm p}+m_{\rm e})/R_{*}\approx GM(m_{\rm p})/R_{*}$, and the thermal 
energy is $2\times\frac{3}{2}kT$, therefore~(\cite{King}), 
\begin{equation}
\label{T1} T_{\rm th}=GMm_{\rm p}/3kR_{*}\approx 5.5 \times 10^{11} \ \ 
K   \ \  ,
\end{equation}
where $R_{*}\sim 1\times 10^{6}$cm is the radius of NS. Replacing it by  
$R_{*}+a\sim 
2\times 10^{9}$cm (recall $P_{\rm b}=1$min corresponds to  $a= 
2\times 10^{9}$cm), we have 
\begin{equation}
\label{T2} T_{\rm th}=GMm_{\rm p}/3k(R_{*}+a)\approx 2\times 10^{8} \ \ K   \ \  .
\end{equation}
This temperature means that the energy released by two distant NSs (in 
a wide binary) decaying to an ultra-compact binary is 
\begin{equation}
\label{DeltaE2} \Delta E_{2}\sim 3 k T_{\rm th} N    \ \  .
\end{equation}
From  Eq.~($\ref{DeltaE1}$) and  Eq.~($\ref{DeltaE2}$) we see that  the two 
energies, $\Delta E_{1}$ and $\Delta E_{2}$, are 
 equivalent  when $\sigma\sim 0.4$. This implies that the 
energy released in the orbital decay can be  consistent with the energy loss 
predicted by the fast cooling of a postsupernova binary. 

The fast  cooling  (on time scales of 10-100 yr) corresponds to a pulsar wind of  
$(10^{-5}M_{\odot}-10^{-6}M_{\odot})$yr$^{-1}$, which interacts 
with the SN envelope~(\cite{Bed01, Ik04}) and causes torques reducing 
both the orbital  angular momentum of the binary system and 
the spin angular momenta of the pulsars. Consequently,  the kinetic 
energy of the binary system is reduced substantially. Thus the 
binary orbit decays from a wide one to an ultra-compact one, and the 
spin frequencies of the two NSs also reduce substantially.   

The above scenario is analogous to the magnetic dipole radiation 
resulting in  the spin-down of a pulsar, assuming that  the energy loss by  
electro-magnetic emission equals the decrease in mechanical  energy 
 corresponding to the spin-down of a NS.

Therefore, an ultra-compact binary can be formed in 10-100yr after SN 
explosion through fast cooling, and the observed SNR age 
can still be greater than  the time scale of the formation. Thus  
the problem in the formation of ultra-compact binaries like 1E1207 
can be naturally explained.


\section{Discussion}

As discussed in Section\ref{sec:disk}, the  enlarged  corotation radius 
of an ultra-compact binary  
makes it possible to accrete also in the  low accretion rate case, i.e., 
$\dot{m}\sim 10^{14}$g s$^{-1}$.  Although this accretion rate is much lower 
than that of normal X-ray binaries, it  can supply  enough falling material to 
produce the observed  cyclotron lines.  
Consequently, the cyclotron features of 
1E1207 should be similar to those of accretion 
pulsars~(\cite{Trumper05,Heindl04}).     

Cold plasma falling freely along the polar magnetic field can be 
stopped above the NS surface by a shock wave formed by radiation 
conduction in a scattering medium~(\cite{Arons91}). 
The in-falling material then forms a hot spot on the surface of NS 
from which X-rays will be radiated. 
The spot has above it a continuum-producing volume through which 
photons must propagate. This creates a 
line forming region, characterized by a magnetic field strength, electron 
(or proton) temperature, and Thompson optical depth~(\cite{Heindl04}).  
The  height of such a volume, $h$,  may influence the X-ray 
luminosity~(\cite{Arons91}),  
\begin{equation}
\label{h} 
L_{x}=\frac{Gm\dot{m}}{R_{*}+h}
\,.
\end{equation}
Considering  Eq.~($\ref{h}$), Eq.~($\ref{rm}$), and assuming a greater accretion rate, 
i.e., $\dot{m}\sim 10^{15}$g s$^{-1}$,  the 
accretion regime is possible ($r_{\rm c}>r_{\rm m}$), even when $B\sim 
10^{14}$G. In other words, the possibility that 1E1207 is a magnetar 
cannot be excluded, and the cyclotron absorption line may 
be caused by protons.

It is clear that the expected line shape  is a strong function of 
the viewing geometry~(\cite{Heindl04}). The variation of the spectrum with the  pulsation 
of the  NS provides evidence of the dependence of the emission  on the viewing angle.

The best fit two-BB model to the  XMM-Newton data indicates an emitting 
radius of $3$km for the soft component, with a temperature of $\sim 
200$eV~(\cite{Bignami03}). 

Since an ultra-compact binary contains two compact stars ( 
separated at $\sim 2\times 10^{9}$cm), which 
cannot be resolved at the distance of $2.1$kpc,   the two BB 
temperatures~(\cite{Mere96, Vas97, Zavlin98, Bignami03}) may be radiation
from the two compact stars. 
 The two temperatures may be caused by the two cooling NSs. 
The accretion induced spots   may also contribute to the nonuniform 
temperature distributions on the NS surfaces, 
reflected in a  pulsation of 1E1207  at the $6\%$ level. 

Therefore, the two BBs should have different pulse frequencies,  
corresponding  to different pulse periods of the two compact stars. This 
prediction  can be checked by   further spectral fitting.

The confrontation of the ultra-compact binary model of 1E1207 with 
the observations of it  can be summarized as follows.

(1) An ultra-compact NS-NS binary system would naturally explain the 
absence of optical counterpart~(\cite{Luca04}). This is now confirmed 
by deep HST observations, with no object in the Chandra error bar 
down to $m_{\rm v}\sim 27$ (Mignani 2006, private communication).

(2)
The  predicted long-periodic effect in 
$\nu$ and $\dot{\nu}$ can explain the observed variation in pulse 
frequency, as well as  in the age and  the  B-field 
discrepancies. Such   phenomena can all be attributed to timing properties 
due to orbital motions.

(3)
An ultra-compact binary makes it possible to enlarge the corotation radius
and hence  accretion may supply the plasma density 
needed for producing  the cyclotron lines in 1E1207.

(4)
The two observed BB temperatures can be interpreted by the cooling temperatures or 
accretion-induced emissions of the two compact stars.

Moreover, the ultra-compact binary scenario predicts the following 
measurable effects:

(a)
The large Doppler shift corresponding to the 
orbital motion of 1E1207 can be directly measured by the new method 
proposed above. 

(b) 
The variation of pulse frequency is not 
only non-monotonic, but also non-periodic (the long-periodic spin-orbit coupling 
effect is not exactly periodic).

(c)
Two BB temperatures should have 
different pulse periods.

\begin{acknowledgements}
 Biping Gong would like to thank Z.Li for useful 
discussion.  
\end{acknowledgements}

{}

\end{document}